\documentclass[%
 reprint,
 aip,
 jcp,
 numerical,
]{revtex4-1}

\usepackage{gensymb}
\usepackage{hyperref}
\usepackage[utf8]{inputenc}
\usepackage{amsfonts,amssymb}
\usepackage{amsmath}
\usepackage{graphicx}
\DeclareMathOperator*{\argmin}{arg\,min}

\newcommand{\mehp}{~m$E_{\rm h}$.{ }}
\newcommand{\meh}{~m$E_{\rm h}${ }}
\newcommand{\mehc}{~m$E_{\rm h}$,{ }}

\begin{document}
\title{Alternative definition of excitation amplitudes in Multi-Reference state-specific Coupled Cluster}

\author{Yann Garniron}%
\affiliation{Laboratoire de Chimie et Physique Quantiques (CNRS 5626), IRSAMC, \\Université P. Sabatier, \\Toulouse (France)}
\author{Emmanuel Giner}%
\affiliation{Dipartimento di Scienze Chimiche e Farmaceutiche, \\Universita di Ferrara, Via Fossato di Mortara 17,\\ I-44121 Ferrara (Italy)}
\affiliation{Max Planck Institut for solid state research, Heisenbergstraße 1, 70569 Stuttgart, Germany}
\author{Jean-Paul Malrieu}%
\affiliation{Laboratoire de Chimie et Physique Quantiques (CNRS 5626), IRSAMC, \\Université P. Sabatier, \\Toulouse (France)}
\author{Anthony Scemama}%
\email{scemama@irsamc.ups-tlse.fr}
\affiliation{Laboratoire de Chimie et Physique Quantiques (CNRS 5626), IRSAMC, \\Université P. Sabatier, \\Toulouse (France)}

\begin{abstract}
A central difficulty of state-specific Multi-Reference Coupled Cluster (MR-CC)
formalisms concerns the definition of the amplitudes of the single and double
excitation operators appearing in the exponential wave operator. If the
reference space is a complete active space (CAS) the number of these amplitudes
is larger than the number of singly and doubly excited determinants on which
one may project the eigenequation, and one must impose additional conditions.
The present work first defines a state-specific reference-independent operator
$\hat{\tilde{T}}^m$ which acting on the CAS component of the wave function $|\Psi_0^m
\rangle$ maximizes the overlap between $(1+\hat{\tilde{T}}^m)|\Psi_0^m \rangle$ and
the eigenvector of the CAS-SD CI matrix $|\Psi_{\rm CAS-SD}^m \rangle$. This
operator may be used to generate approximate coefficients of the Triples and
Quadruples, and a dressing of the CAS-SD CI matrix, according to the
intermediate Hamiltonian formalism. The process may be iterated to
convergence. As a refinement towards a strict Coupled Cluster
formalism, one may exploit reference-independent amplitudes
provided by $(1+\hat{\tilde{T}}^m)|\Psi_0^m \rangle$  to define a reference-dependent
operator $\hat{T}^m$ by fitting the eigenvector of the (dressed) CAS-SD CI matrix.
The two variants, which are internally uncontracted, give rather similar
results. The new MR-CC version has been tested on the ground state potential
energy curves of 6 molecules (up to triple-bond breaking) and a two excited
states. 
The non-parallelism error with respect to the Full-CI curves is of the order of
1\mehp
\end{abstract}

\maketitle

\section{Introduction}
The single-reference Coupled Cluster (CC)
formalism\cite{CC_Nucl1,CC_Nucl2,CC_Quantum,Barlett_et_al_1990} is the standard
technique in the
study of the ground state of closed-shell molecules, i.e. those for which a
mean-field treatment provides a reasonable zero-order single-determinant
wave-function $\Phi_0$. This method incorporates the leading contributions to
the correlation energy in a given basis set, it is based on the linked-cluster
theorem\cite{Goldstone} and is size-consistent since it is free from unlinked contributions.
The method generates an approximate wave function under the action of a wave
operator $\hat{\Omega}$ acting on the single-determinant reference $\Phi_0$,
and assumes an exponential character to the wave operator
\begin{equation}
\Psi = \hat{\Omega} \Phi_0 = \exp \left( \hat{T} \Phi_0 \right)
\end{equation}
The most popular version only introduces single and double excitation operators
in $\hat{T}$, and is known as the Coupled Cluster Singles and Doubles (CCSD)
approximation. It incorporates the fourth order correction of the quadruply
excited determinants. The lacking fourth order
contribution concerns the triply excited determinants, which may be added in a
perturbative manner. The CC equations, obtained by projecting the eigenequation
on each of the Singles and Doubles lead to coupled quadratic equations. In
practice, guess values of the amplitudes of the $\hat{T}_{0\rightarrow i}$ operators
appearing in the $\hat{T}$ operator may be taken as the coefficients of the
Singles and Doubles $|i\rangle$ in the intermediate normalization of the SD CI
vector. The solution of the CC equations may be obtained by treating the effect
of the Triples and Quadruples as an iterative dressing of the SD CI matrix,\cite{heff_ccsd}
according to the Intermediate Effective Hamiltonian (IEH)
theory.\cite{Malrieu_Hamiltonien,Kirtman_1981} The field of
application of this method, which is both intellectually satisfying and
numerically efficient, is however limited to the systems and the situations
where a single-determinant zero-order description is relevant. This is
no longer the case when chemical bonds are broken, creating open shells, as
occurs in most of the chemical reactions. The magnetic systems generally
present several open shells and the low spin-multiplicity states are intrinsically
of multiple-determinant character. Due to near degeneracies, most of the
excited states are not only of multi-determinantal but of multi-configurational
character.  The conception of a multi-reference counterpart of the CCSD
formalism is highly desirable, and has been the subject of intense
research. The most comprehensive review has been given by R. Bartlett and his
colleagues.\cite{Lyakh_2012} For formal reasons and in particular to treat
correctly the breaking of bonds, the reference space, or model space, is usually
taken as a Complete Active Space (CAS), i.e. the Full-CI of a well-defined
number of electrons (the active electrons) in a well-defined set of orbitals
(the active MOs). The other MOs are called inactive. Let us label $|I\rangle,
|J\rangle, \dots$ the reference determinants. The determinants $|i\rangle,
|j\rangle, \dots$ which interact with the reference space are obtained under
purely inactive or semi-active single and double excitations, they generate the
CAS-SD CI space, the diagonalization of which provides a size-inconsistent
energy $E_{\rm CAS-SD}^m$ and the corresponding eigenvector,
\begin{eqnarray}
|\Psi_{\rm CAS-SD}^m\rangle
&=& |\Psi_{0}^m\rangle + |\Psi_{\rm SD}^m\rangle  \nonumber\\
&=& \sum_{I\in {\rm CAS}} C_I^m + |I\rangle + \sum_{i\notin {\rm CAS}} c_i^m |i\rangle 
\end{eqnarray}
with $\langle \Psi_{\rm CAS-SD}^m| \Psi_{\rm CAS-SD}^m \rangle = 1$.

One strategy, which is not very aesthetic since it breaks the symmetry between
degenerate reference determinants, but which has given rather satisfactory
results, consists in selecting (eventually in an arbitrary manner) a specific
single reference and in introducing in the wave operator the multiple
excitations which generate the other references (the other determinants of the
model space). The other strategies consider all the references on an equal
footing, and are really multi-reference. Let us call $N$ the number of
references, and $n$ the number of SD determinants. If the treatment pretended to
provide $N$ eigenvectors simultaneously, one might define the $N\times n$ amplitudes
sending from the references to the outer-space determinants, in a unique manner
but this state-universal approach is not practicable when the model
space is a CAS.

Most of the proposed formalisms are state-specific. In this
case one faces the famous multi-parentage problem. This problem is recalled in
section 2A.  Sufficiency conditions have to be imposed.\cite{Mahapatra_et_al_1998}
One solution was proposed by Mukherjee and coworkers, and has been widely
tested.\cite{Das_2010,Szabados_2011,Mahapatra_1999} Another one
had been proposed earlier by one of us (JPM) and coworkers.\cite{Meller_1996}
It consists, for a
given outer-space determinant, in scaling the amplitudes of the various
excitation operators $\hat{T}_{I\rightarrow i}$ on the interaction between the
outer-space determinant and its parents. A recent work has implemented this
second solution of the state-specific MR-CC problem and has tested its accuracy
and robustness on a series of molecular benchmarks, comparing its results to
the Full-CI (FCI) energies.\cite{Giner_2016} The present work proposes an
alternative process to define the amplitudes of the excitation operators.
The state-specific MR-CC formalisms are usually based on the
Jeziorski-Monkhorst\cite{Jeziorski_1981} splitting of the wave operator into a sum of
operators acting individually on the various references
\begin{equation}
\hat{T}^m = \sum_I \hat{T}_I^m |I \rangle \langle I|
\label{eq:jm}
\end{equation}

We shall leave in a first time this assumption and define in
section~\ref{section2b} a reference-independent operator $\hat{T}$ which acting
on the component of the desired state in the model space, $|\Psi_0^m\rangle$,
provides a vector as close as possible to the CAS-SD eigenvector. This solution,
defining reference-independent amplitudes of the excitations, may be exploited
directly to generate approximate values of the coefficients of the triply and
quadruply excited determinants, according to the exponential structure of the
wave operator. From these coefficients one may dress the CAS-SD CI matrix,
redefine amplitudes and iterate the process to convergence. This solution,
presented in section 2C, is not an MR-CC technique, one may call it an
exponential dressing of the CAS-SD CI matrix. Section 2D redefines
reference-dependent excitation amplitudes from the reference-independent
amplitudes by a fitting of the previous amplitudes on the coefficients of the
Singles and Doubles of the (dressed) CAS-SD CI eigenvector. This represents an
alternative solution to multi-parentage problem and opens the way to a strict
MR-CC formalism. Section 3 presents a series of numerical tests on the bond
breaking of single, double and triple bonds in ground states of molecules as
well as a few tests on excited states.  The results are compared to our
previous proposal and with full Configuration Interaction (FCI) results.

\section{Formalisms}

In this section, all the presented formalisms are state-specific. To simplify the notations we will consider that
the state superscript $m$ is implicit for the wave functions ($\Psi^m \rightarrow \Psi$) and for the excitation operators
($\hat{T}^m \rightarrow \hat{T}$).

\subsection{The multi-parentage problem in the Jeziorski-Monkhorst approach}
\label{section2a}

Since one wants to produce a MR-CCSD method, one may start from a preliminary
CAS-SD CI calculation which will help to fix guess values of the amplitudes of
the excitation operators. Let us call $|I\rangle, |J\rangle, \dots$ the
determinants of the CAS, i.e. the so-called reference vectors, and $|i\rangle,
|j\rangle, \dots$ the Singles and Doubles which do not belong to the CAS and
interact with them. The resulting approximate wave function of the
targeted state $|\Psi\rangle$ is written
\begin{equation}
|\Psi_{\rm CAS-SD} \rangle = \sum_{I} C_I |I\rangle + \sum_{i} c_i |i\rangle
\end{equation}
Although this function is not size consistent one may note that the coefficients
on the CAS determinants are no longer those of the CAS-CI~: they incorporate the
effect of the dynamical correlation on the composition of the CAS component of
the wave function.

In CC formalisms the wave operator $\hat{\Omega}$ is assumed
to take an exponential form
\begin{equation}
 \hat{\Omega} = \exp (\hat{T})
\end{equation}
and in our previous MR-CC formalism\cite{Giner_2016} the Jeziorski-Monkhorst
structure of the wave operator was adopted, introducing reference-specific wave
operators acting specifically on each reference vector (Eq.\ref{eq:jm}).
One may exploit the knowledge of the CAS-SD CI eigenvector to determine guess
operators $\hat{T}_I$ defined in such a manner that
\begin{equation}
|\Psi_{\rm CAS-SD} \rangle = \sum_I C_I\, \hat{T}_I |I\rangle
\label{eq:cassd}
\end{equation}
The $\hat{T}_I$ operators are a sum of single and double excitations $\hat{T}_{I \rightarrow i}$
possible on $|I\rangle$, multiplied by an amplitude $t_{I\rightarrow i}$
\begin{equation}
\hat{T}_I = \sum_i t_{I\rightarrow i} \hat{T}_{I\rightarrow i}
\end{equation}
In the single-reference CC the amplitudes of the excitation operators are
obtained by projecting the eigenequation on the singly and doubly excited
determinants, the number of unknowns is equal to the number of equations. This
is no longer the case in the MR context~: projecting the eigenequation on each
on the singly or doubly excited vectors $|i\rangle$ is not sufficient to
define the amplitudes $t_{I\rightarrow i}$ since for many classes of
excitation an outer-space determinant interacts with several references,
$|i\rangle = \hat{T}_{I\rightarrow i} |I\rangle = \hat{T}_{J\rightarrow i} |J\rangle$.
The condition 
\begin{equation}
C_i = \sum_I t_{I\rightarrow i}\, C_I
\end{equation}
is not sufficient to define the amplitudes, even if one restricts the excitation
operators to single and double excitations. Additional constraints have to be
introduced to fix the amplitudes, and this is the famous \emph{multi-parentage problem}.
The number of amplitudes is larger than the number of outer-space
determinants so that one cannot determine directly guess values of the amplitudes
from Eq.~\ref{eq:cassd}. Different additional constraints have been proposed.
One of them consists in scaling the amplitudes on the Hamiltonian interactions
between the references and the outer space determinants, 
\begin{equation}
\frac{ t_{I\rightarrow i} }{  t_{J\rightarrow i} } = \frac{ \langle i | \hat{H} | I \rangle }{ \langle i | \hat{H} | J \rangle }.
\end{equation}
This constraint is expressed as
\begin{equation}
t_{I\rightarrow i} = \lambda_i \langle i | \hat{H} | I \rangle
\end{equation}
where 
\begin{equation}
\lambda_i = \frac{c_i}{\langle i |\hat{H} | \Psi_0 \rangle}
\label{eq:lambda}
\end{equation}

This solution has been recently implemented\cite{Giner_2016} and shown to provide excellent
agreements with Full-CI results on a series of molecular problems. It only
presents minor stability problems in comparison with the present suggestion when
the term ${\langle i |\hat{H} | \Psi_0 \rangle}$ is small.
From now on, we will refer to this method as the $\lambda$-MR-CCSD.

\subsection{Introduction of reference-independent amplitudes}
\label{section2b}

The present formalism will leave in the first step the Jeziorski-Monkhorst formulation of the wave operator
and will consider the possibility to define a unique state-specific reference-independent operator $\hat{T}$, written
as a sum of single and double excitation operators,
\begin{eqnarray}
\hat{T}
&= &\sum_{mnpq} t_{mn \rightarrow pq}\, a^\dagger_p a^\dagger_q a_n a_m +
  \sum_{mp} t_{m \rightarrow p}\, a^\dagger_p a_m \\
& = & \sum_{mnpq} t_{mn\rightarrow pq}\, \hat{T}_{mn \rightarrow pq} + 
  \sum_{mp} t_{m \rightarrow p}\, \hat{T}_{m\rightarrow p}
\end{eqnarray}
where the indices $p$ and $q$ run on the virtual and active MOs and the indices $m$ and $n$ run on the
inactive occupied and active MOs, excluding the possible occurrence of 4 active MOs.

%At this stage, introducing single excitation operators in $\hat{T}$ does not seem crucial since the
%generalized Brillouin's theorem
%\begin{equation}
% \langle \hat{T}_{m \rightarrow p} \Psi_0 | \hat{H} | \Psi_0 \rangle = 0 \; \forall m,p.
%\end{equation}
%is fulfilled when using CASSCF molecular orbitals.
%The so-generated vectors should have weak components in the wave function. For simplicity
%we present the method considering only the double excitation operators, but all
%the applications presented in the next sections also include the single
%excitation operators.
This operator has the same form as the one introduced by the internally-contracted
MR-CC (ic-MRCC) method by Evangelista and Gauss,\cite{Evangelista_2011} and by
Hanauer and K\"ohn,\cite{Hanauer_2012} but it differs by both its determination and by
the way we use it, as will appear later. The ic-MRCC method determines the amplitudes of
the excitation by solving the projected Coupled Cluster equations, where the amplitudes
appear as linear and quadratic terms. Hereafter we exploit the knowledge of the CAS-SD
CI eigenvector to determine guess values of the reference-independent amplitudes. These
excitation amplitudes will be used later on to estimate the coefficients of the
Triples and Quadruples, and perform an iterative dressing of the CAS-SD CI
matrix introducing the coupling between the Singles and Doubles with the
Triples and Quadruples.

We propose a criterion to fix the amplitudes $ t = \{t_{mn\rightarrow pq}, t_{m \rightarrow p}\}$.
Given the fact that we have at our disposal the CAS-SD wave function, a natural
way to solve this overdetermined problem is to minimize the distance between
the CAS-SD vector and the vector obtained by applying the $(1+\hat{T})$ operator
on the CAS wave function
\begin{equation}
\begin{split}
\argmin_{t} \Vert (1+\hat{T}) | \Psi_0\rangle - | \Psi_{\rm CAS-SD} \rangle \Vert \\ 
= \argmin_{t} \Vert \hat{T} | \Psi_0\rangle - | \Psi_{\rm SD} \rangle \Vert,
\end{split}
\end{equation}
$\hat{T} | \Psi_0\rangle$ being normalized such that $\Vert \hat{T} | \Psi_0 \rangle
\Vert = \Vert |\Psi_{\rm SD}  \rangle\Vert$.

To perform the minimization, we build the $N_{\rm SD} \times N_{\rm t}$ transformation matrix
$A_{i, mn\rightarrow pq} = \langle i | \hat{T}_{mn\rightarrow pq} | \Psi_0 \rangle$
 which maps from the outer space of determinants $\{|i\rangle\}$ to the
space of excited wave functions $\{\hat{T}_{mn\rightarrow pq}|\Psi_0\rangle\}$,
and we search for the vector of amplitudes $\bf{t}$ which minimizes
$\Vert {\bf A}.{\bf t} - {\bf c} \Vert$ by solving the normal equations
\begin{equation}
({\bf A}^\dagger {\bf A}) {\bf t} = {\bf A}^\dagger {\bf c}
\label{eq:normal}
\end{equation}
Note that in the single-reference case, ${\bf A}$ is a permutation matrix and the
CAS-SD wave function is exactly recovered.

The matrix ${\bf A}$ is usually so large that the use of standard singular
value decomposition (SVD) routines to obtain the least squares solution is
prohibitive.

Let us first consider the most numerous 2-hole-2-particle inactive double excitations
$\hat{T}_{jk\rightarrow rs}$. Acting on a determinant $|I\rangle$ the operator creates a determinant
$|i\rangle = \hat{T}_{jk\rightarrow rs} | I \rangle$ which
can only be produced by this process. Therefore, the corresponding rows of ${\bf A}$
contain only one non-zero element located in the $jk\rightarrow rs$ column with value 
$A_{i, jk\rightarrow rs} = C_I$.
The condition fixing the amplitude $t_{jk\rightarrow rs}$ is given by
\begin{equation}
 \argmin_{t_{jk\rightarrow rs}}  \Vert \hat{T}_{jk\rightarrow rs} |\Psi_0 \rangle
 t_{jk\rightarrow rs} - |\Psi_{\rm SD} \rangle \Vert \\
\end{equation}
which is obtained by minimizing
\begin{equation}
 \min_{ t_{jk\rightarrow rs } } \left( \sum_I C_I \, t_{jk\rightarrow rs} - \sum_i c_i \langle i | \hat{T}_{jk\rightarrow rs}|I\rangle \right)^2 \\
\end{equation}
This condition turns out to be satisfied for
\begin{equation}
 t_{jk\rightarrow rs} = \frac{\sum_I C_I \, c_i }{\sum_I C_I^2}
\end{equation}
One may notice that this is the weighted average of the ratios between the coefficients
of the doubly excited determinants $|i\rangle$ and the
coefficient of their unique reference generator,
\begin{equation}
 t_{jk\rightarrow rs} = \frac{1}{\sum_I C_I^2} \left( \sum_I C_I^2 \left(\frac{c_i}{C_I}\right) \right)
\end{equation}

For all the remaining active excitations, {\bf A} remains sparse since the maximum
number of non-zero elements per column is equal to the number of reference
determinants. Hence, we use Richardson's iterative procedure\cite{Richardson_1911}
\begin{equation} 
\left\{ \begin{array}{l}
{\bf t}_0 = {{\bf A}}^{\dagger}{\bf c} \\
{\bf t}_{n+1} = {{\bf A}}^{\dagger}{\bf c} + \left({\bf I} - {{\bf A}}^{\dagger}{\bf A} \right){\bf t}_n \\
\end{array} \right.
\label {eq:iterative}
\end{equation} 
which may be implemented very efficiently using sparse matrix products.

There are cases where multiple amplitudes applied to different references lead to same determinant : 
$\hat{T}_{jk\rightarrow rs} |I \rangle = \hat{T}_{lm\rightarrow tv} |J \rangle = |i\rangle$. If this determinant
can be reached by no other process, there is an infinity of solutions for the amplitudes. 
The solution of Eq.(\ref{eq:normal}) is ${\bf A}^+ {\bf c}$
where ${\bf A}^+$ is the pseudo-inverse of ${\bf A}$, so the solution
obtained minimizes the norm of the amplitude vector.\cite{Golub_1965}
In this way, the arbitrariness brought by the null space of ${\bf A}$ is
minimized in the amplitude vector, and one obtains the most sensible solution.

%Notice that the ic-MRCC formalisms also employ the SVD procedure. but only to
%remove the linear dependences in the basis of the contracted excited wave
%functions. Here, we use the SVD for the determination of the amplitudes
%maximizing the overlap between the CAS-SD CI vector and
%$(1+\hat{T})|\Psi_0\rangle$.

\subsection{Evaluation of the coefficients of Triples and Quadruples and iterative dressing of the CAS-SD CI matrix}
\label{section2c}

The so-determined excitation operator $\hat{T}$  may be used to generate
approximate values of the coefficient of the Triples and Quadruples as obtained
by the action of $\frac{1}{2}\hat{T}^2$. Actually one may assume, in the spirit of
the internally-contracted MR-CC methods, that the wave operator $\hat{\Omega}$ generating the
correlated wave function $\Psi$ from $\Psi_0$,
\begin{equation}
\Psi = \hat{\Omega} \Psi_0
\end{equation}
has an exponential structure,
\begin{equation}
\hat{\Omega} = \exp (\hat{T})
\end{equation}
But this form will be simply used to estimate the coefficients of the triply and quadruply
excited determinants $\{|\alpha\rangle \}$, leaving the internally-contracted structure of the
outer-space. The coefficients of these  determinants are estimated as
\begin{equation}
c_\alpha = \frac{1}{2} \langle \alpha | \hat{T}^2 | \Psi_0 \rangle.
\end{equation}

In practice all the determinants $\{|\alpha \rangle \}$ are generated. For each
$|\alpha\rangle$ one finds the reference determinants $\{|I\rangle\}_{\alpha}$ 
which differ by at most 4 orbital substitutions from $|\alpha \rangle$
(its grand-parents). One then identifies the set of
complementary excitations $\left\{ (p, q)\, \big|\, \hat{T}_p \hat{T}_q |I\rangle = |\alpha \rangle \right\}$
as the products which
generate $|\alpha \rangle$ from $\{|I\rangle\}_{\alpha}$.
The genealogy of $|\alpha \rangle$
contributes to its associated coefficient $c_{\alpha}$ by the quantity $t_p t_q C_I$.
Knowing $|\alpha\rangle$, one also knows the Singles and Doubles $\{|i\rangle\}_{\alpha}$ with which
it interacts through the matrix elements $\langle i | \hat{H} | \alpha \rangle$, and in the
eigenequation relative to $|i\rangle$
\begin{equation}
\begin{split}
\left( \langle i | \hat{H} | i \rangle - E \right) c_i + 
\sum_J \langle i | \hat{H} | J \rangle C_J + \\
\sum_j \langle i | \hat{H} | j \rangle c_j + 
\sum_\alpha \langle i | \hat{H} | \alpha \rangle c_\alpha = 0 \;.
\end{split}
\end{equation}
One may replace the last sum by a dressing of the matrix elements between the determinant
$|i\rangle$ and the references which are grand-parents of $|\alpha\rangle$,
\begin{equation}
\langle i | \Delta | I \rangle = \sum_\alpha \langle i | \hat{H} | \alpha \rangle
\left(
\sum_{\left\{(p,q) \big| \hat{T}_p \hat{T}_q |I\rangle = |\alpha \rangle \right\}} t_p t_q \right)
\end{equation}
since 
\begin{equation}
\sum_I \langle i | \Delta | I \rangle C_I = \sum_\alpha \langle i | \hat{H} | \alpha \rangle c_\alpha.
\end{equation}

The effect of the Triples and Quadruples is incorporated as a change of the
columns of the CAS-SD CI matrix concerning the interaction between the
references and the Singles and Doubles. This type of dressing was already
employed in our previous MR-CC implementation.\cite{Giner_2016}
One will find in the same
reference the practical procedure to make the dressed matrix Hermitian without
any loss of information. Of course the whole process may be iterated. The
diagonalization of the dressed CAS-SD CI matrix provides new values of the
coefficients, not only of the Singles and Doubles which no longer suffer from
the truncation, but also those of the references~: the method is fully
non-contracted. From the new wave function new amplitudes are obtained, a new
dressing is defined and the process is repeated till convergence, which is
usually rapidly obtained (3-4 iterations).

This formalism is not a strict MR-CC method since we exploit the CAS-SD CI
function, and since this function slightly differs from the vector resulting
from the action of $\hat{T}$ on the vector. Although the distance between these two
vectors has been minimized they are not identical,
$(1+\hat{T}) |\Psi_0 \rangle \ne |\tilde{\Psi}_{\rm CAS-SD}\rangle$.

Once the $\hat{T}$ operator has been obtained one might imagine a contracted
exponential formalism calculating $\hat{T}^2 |\Psi_0 \rangle$ and the
interaction between $\hat{T} |\Psi_0 \rangle$ and $\hat{T}^2 |\Psi_0 \rangle$,
but this calculation requires to return to the determinants. This formalism
would remain internally contracted and would be less accurate than the
procedure we propose. Actually in this version the deviations of the
approximate reference-independent amplitudes from optimal ones, those which
would generate the exact coefficients of the Singles and Doubles, only affects
the evaluation of the coefficients of the Triples and Quadruples, and these
deviations represent a minor source of error in the correction restoring the
size extensivity. This reliability  will be illustrated in the numerical tests.

\subsection{State-specific MR-CC variant}
\label{section2d}

% Donner les bornes des mu_i er discuter que ce sacrifice des valeurs tres
% grandes n'a pas d'effet dramatique, mair reconnaitre que ca peut introduire
% un peu de bruit.  Revenir sur le fait qu'on fait bien du Jez-Monk.

In order to return to a MR-CC formalism, one may simply exploit the
reference-independent amplitudes as an initial guess to define reference-dependent
amplitudes. Currently the determinant $|i\rangle$ belonging to the
Singles and Doubles has a coefficient $\tilde{c}_i$ in $\hat{T}|\Psi_0\rangle$
\begin{equation}
\tilde{c}_i = \langle i | \hat{T} | \Psi_0 \rangle = \sum_{ \left\{(I,l) \big| \hat{T}_l|I\rangle = |i\rangle \right\}} t_l \, C_I
\end{equation}
which differs from the coefficient $c_i$ in $|\Psi_{\rm SD}\rangle$.
One can define a parameter $\mu_i$, specific of the determinant $|i\rangle$,
\begin{equation}
\mu_i = \frac{c_i}{\tilde{c}_i} 
\label{eq:mu}
\end{equation}
which multiplying $\tilde{c}_i$ will produce the exact coefficient $c_i$ of
$|i\rangle$ in the (dressed) CAS-SD CI eigenvector. 
So the previous reference-independent amplitudes have now become reference-dependent.
The excitation
$\hat{T}_l$ which excites $|I\rangle$ to $|i\rangle$ ($|i\rangle = \hat{T}_l |I\rangle$)
receives a reference-dependent amplitude
\begin{equation}
t_{I\rightarrow i} = t_{l,I} = \mu_i t_l \; .
\end{equation}
The same excitation will receive a somewhat different amplitude when it acts on another reference
$t_{l,J} \ne t_{l,I}$.

As the overlap between $(1+\hat{T})|\Psi_0\rangle$ and
$|\Psi_{\rm CAS-SD}\rangle$ has been maximized the coefficients $\tilde{c}_i$
and $c_i$ are expected to be very close in particular if $c_i$ is large, and the
parameter $\mu_i$ should be close to 1, at least for the determinants which
contribute significantly to the wave function.
In practice we observe this tendency, but the smallest coefficients are sacrificed 
during the maximization of the overlap and their $\mu_i$ can be very far from 1.
This introduces some instabilities in the iterations, so we chose to limit
the values of $\mu_i$ in the $[-2,2]$ range. The effect on the stability of
the iterations is significant, and the effect on the energy is not noticeable. 

This version returns to the Jeziorski-Monkhorst formalism as the wave operator
again is a sum of reference-specific operators. The so-obtained amplitudes may
be exploited to generate the coefficients of the Triples and Quadruples, and one
may follow the same strategy as in our previous formalism, with an iterative
column dressing of the interactions between the Singles and Doubles and the
references. In a strict Coupled Cluster formalism, one should redefine the
amplitudes of the double excitations by subtracting the products of the
complementary single excitations they contain. For the sake of simplicity, as we
did in our previous work, we did not proceed to this revision, the difference
concerning only fifth-order perturbative corrections.

In what follows, we will refer to this method as $\mu$-MR-CCSD as it involves the
$\mu_i$ (Eq.\ref{eq:mu}).

\section{Numerical tests}
\label{section3a}

In this section, we compare the here-proposed dressed CAS-SD and MR-CCSD to the
MR-CCSD presented in ref~\cite{Giner_2016} on standard benchmark
systems.\cite{Engels_Putzka_2009,Purvis_1982,Purvis_1983,Laidig_1984,Mahapatra_1999,Kallay_2002,Hanrath_2005,Evangelista_2006,Evangelista_2010,Hanauer_2011,Aoto_2016,Hanauer_2012,Das_2010,Evangelista_2011}
To differentiate those two variants, we will label $\lambda$-MR-CCSD the
variant of ref~\cite{Giner_2016} (Eq.~\ref{eq:lambda}) and the MR-CCSD of this
work will be labeled $\mu$-MR-CCSD (Eq.~\ref{eq:mu}).

The basis set used is Dunning's cc-pVDZ,\cite{Dunning_1989}
and the molecular orbitals were obtained using the CAS-SCF code present in
GAMESS.\cite{gamess} All the following calculations were made using the Quantum
Package,\cite{qp} an open-source library developed in our group. Full-CI
energies were obtained using the CIPSI algorithm,\cite{huron_jcp_1973,cipsi2,Giner_2013}
and the accuracy of the total energies is estimated to be of the order of
$10^{-5} E_{\rm h}$.  In all the calculations (Full-CI, CAS-SD and MR-CC), only the
valence electrons are correlated (frozen core approximation).

%\subsection{Quality of the fit of the CAS-SD eigenvector through reference-independent amplitudes}
%\label{section3a}
%
%\begin{figure}
%\begin{center}
%\includegraphics[angle=270,width=0.9\columnwidth]{fit.eps}
%\end{center}
%\caption{Dissociation of N$_2$. Energy difference $E[(1+\hat{T})\Psi_0] -
%E[\Psi_{\rm CAS-SD}]$ in atomic units (purple curve) and overlap $\langle
%(1+\hat{T})\Psi_0|\Psi_{\rm CAS-SD}\rangle$ (green curve).}
%\label{fig:fit}
%\end{figure}
%To illustrate the quality of the obtained amplitudes, we present in
%figure~\ref{fig:fit} the overlap $\langle (1+\hat{T})\Psi_0|\Psi_{\rm
%CAS-SD}\rangle$ along the dissociation curve of N$_2$, as well as the difference
%of the energies obtained by both wave functions. The CAS is made of 6 electrons
%in 6 MOs.
%The overlap between the two wave functions is always greater than 0.999 all along the
%curve.  The quality of the fit is better at the equilibrium than at the
%dissociation, and the maximum loss in correlation energy is 4.6 milliHartree,
%which represents 2\% of the correlation energy gained by the inclusion of the
%Singles and Doubles on the CAS wave function.

\subsection{Bond breaking}
\label{section3b}

\begin{figure*}
\begin{center}
\includegraphics[angle=270,width=0.4\textwidth]{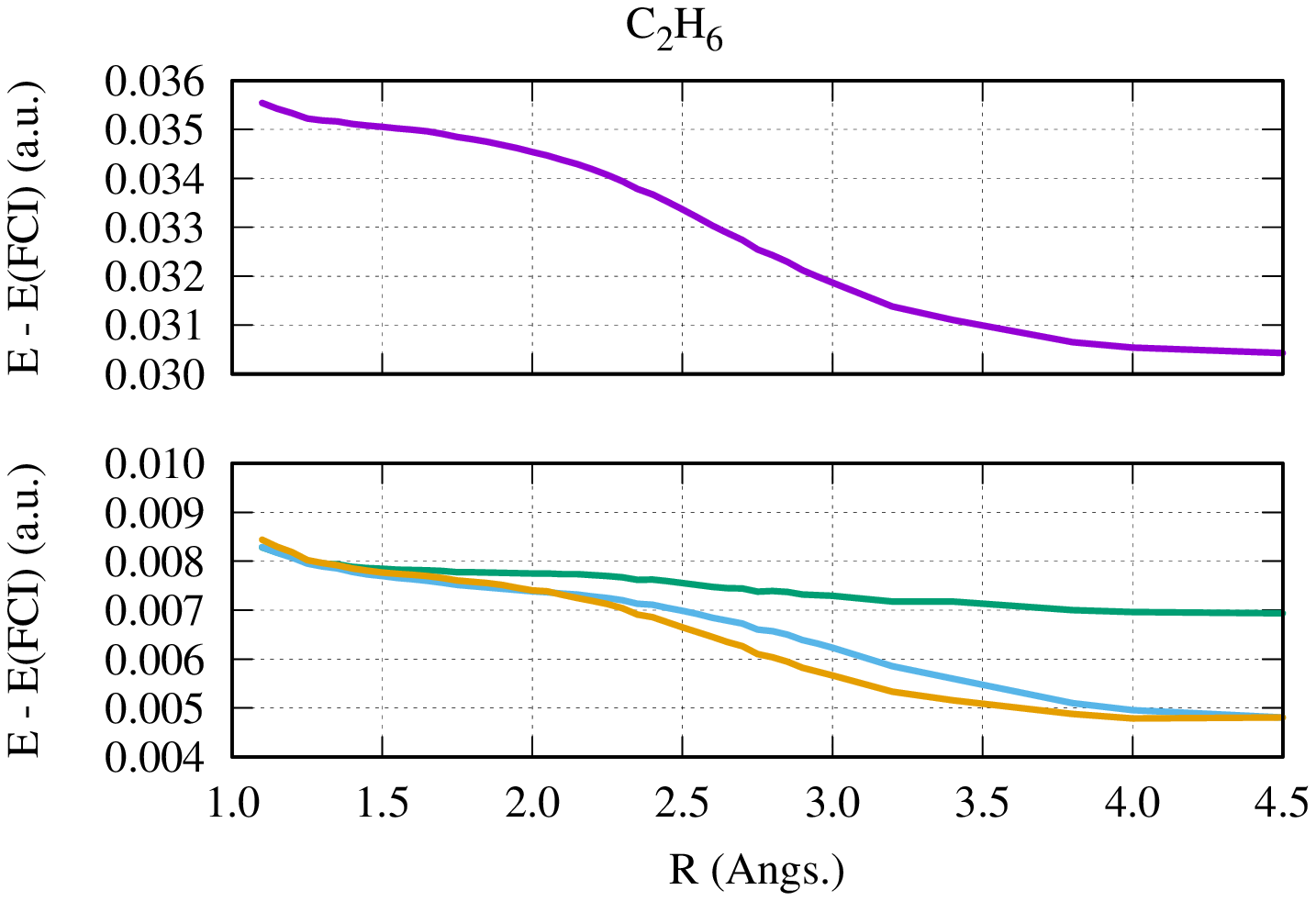}
\includegraphics[angle=270,width=0.4\textwidth]{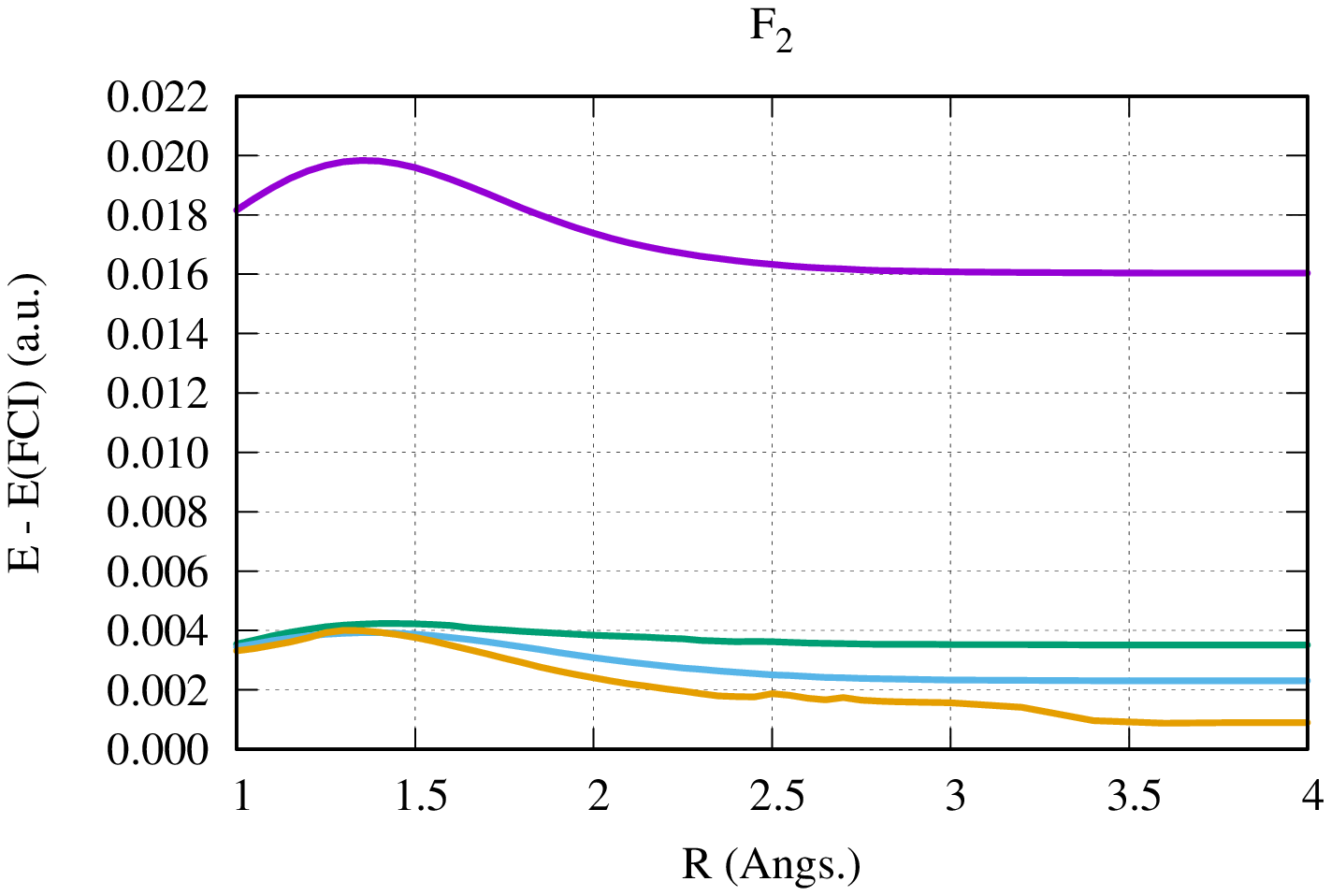}
\includegraphics[angle=270,width=0.4\textwidth]{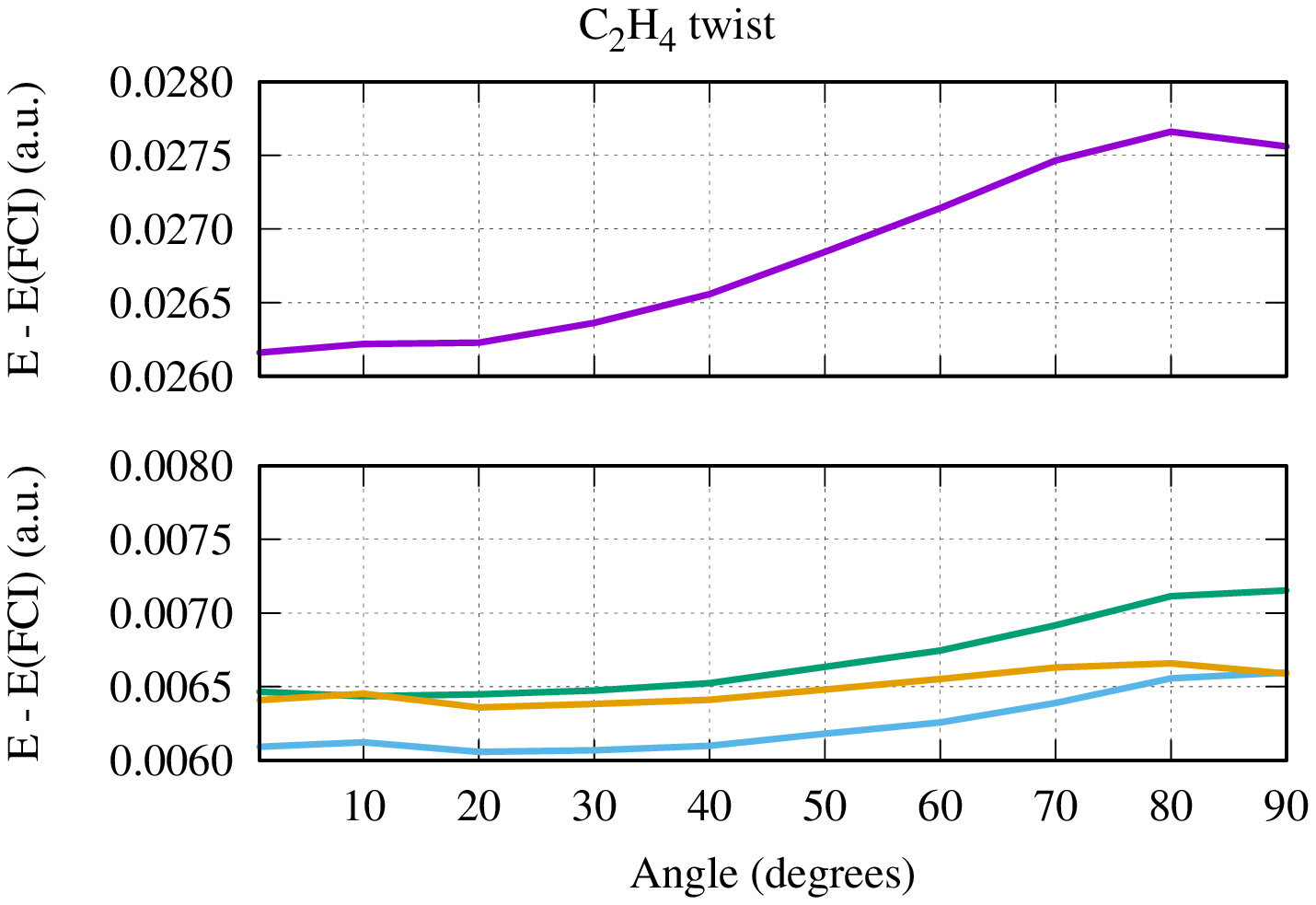}
\includegraphics[angle=270,width=0.4\textwidth]{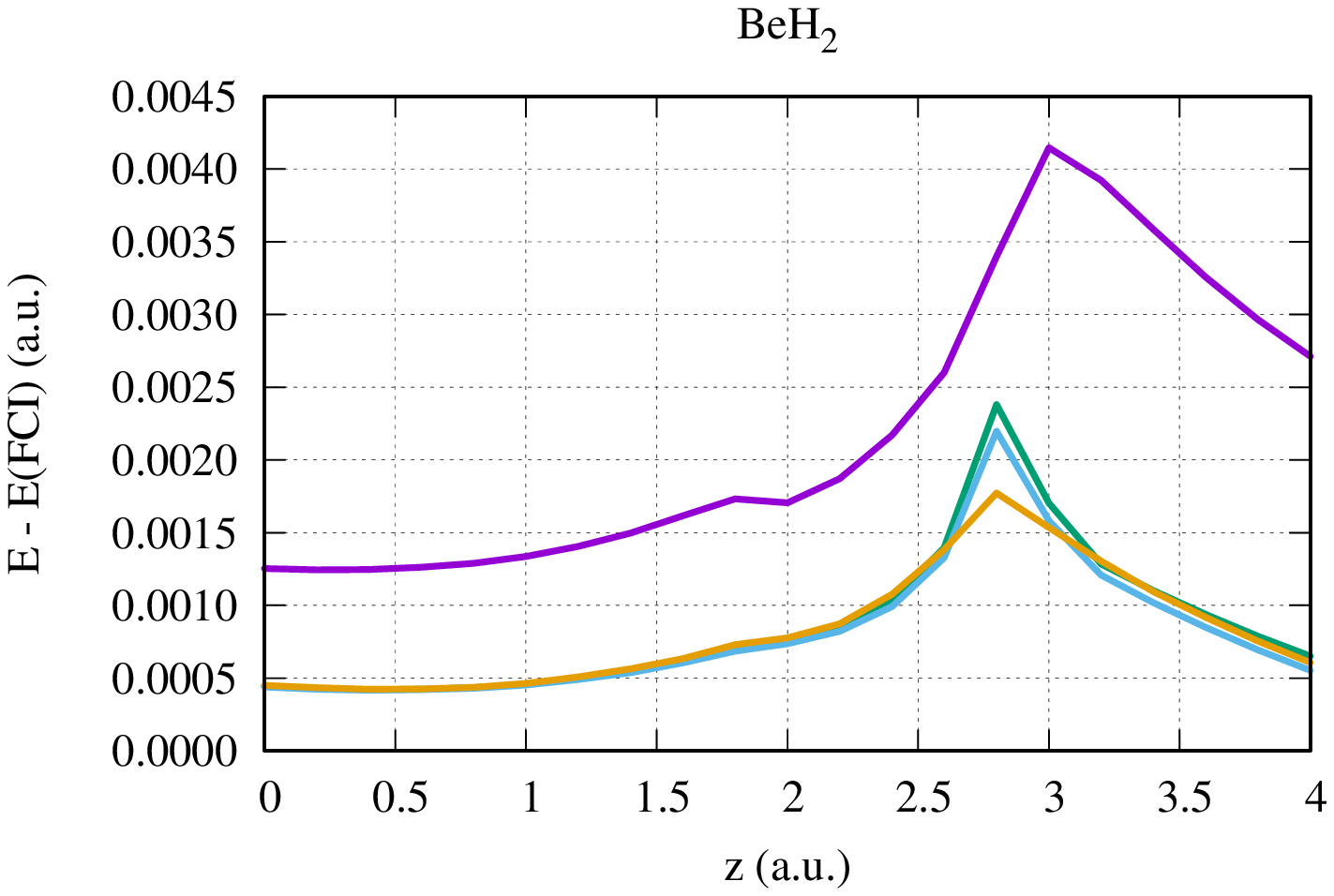}
\includegraphics[angle=270,width=0.4\textwidth]{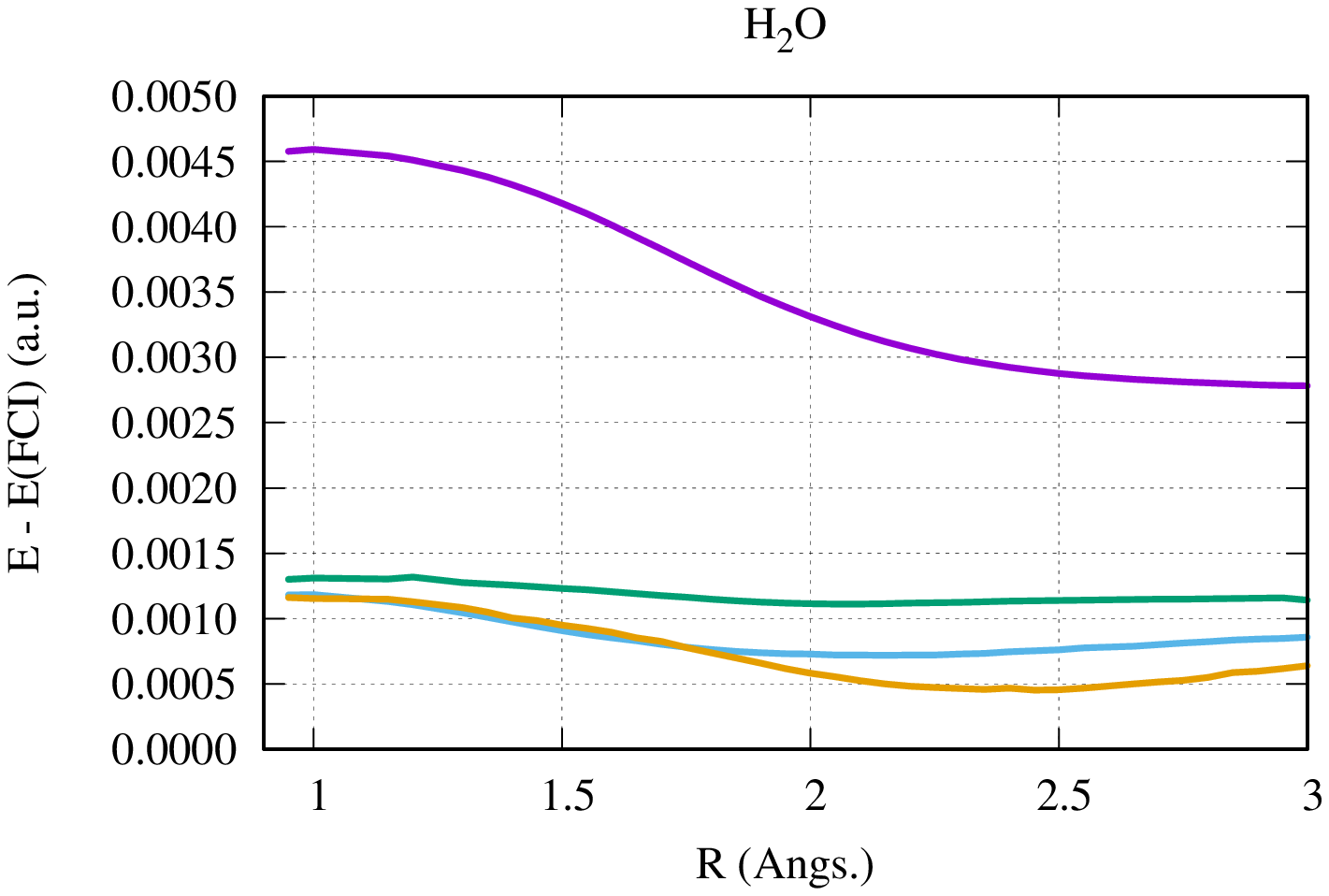}
\includegraphics[angle=270,width=0.4\textwidth]{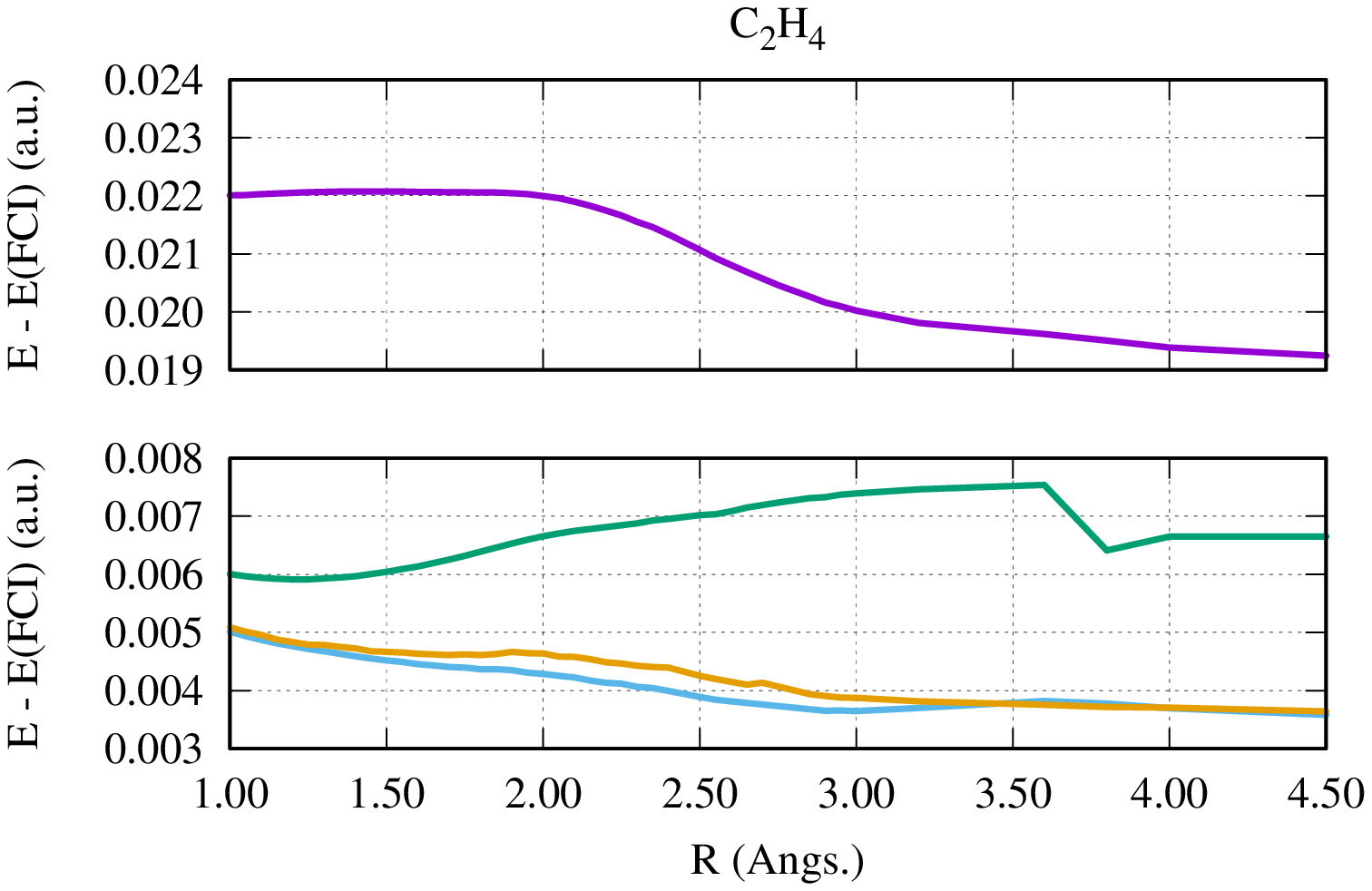} 
\includegraphics[angle=270,width=0.6\textwidth]{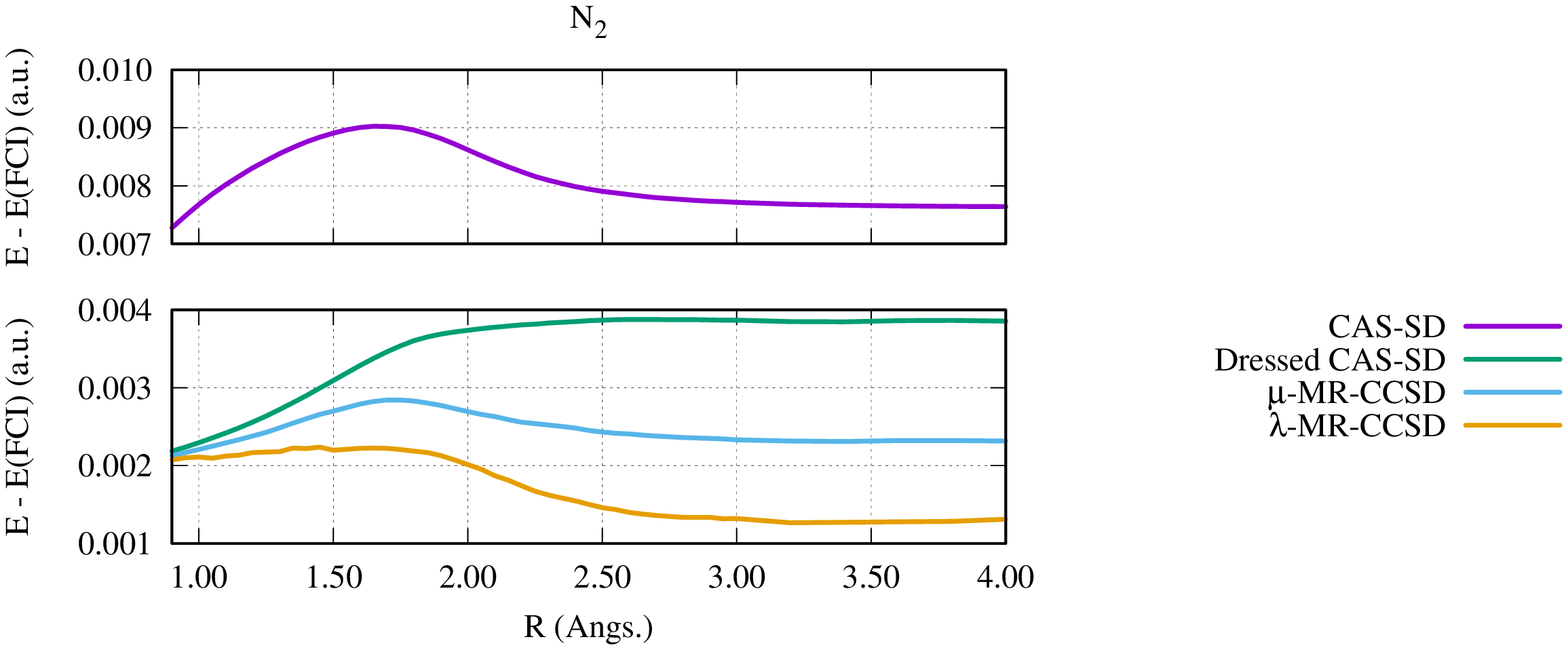}
\end{center}
\caption{Dissociation curves. Difference with respect to the Full-CI energy using
the MR-CCSD method presented in~ref\cite{Giner_2016} and with the MR-CCSD method proposed
in this work, as well as the CAS-SD and the dressed CAS-SD.}
\label{fig:npe}
\end{figure*}

For all the applications we compare the dressed CAS-SD and $\mu$-MR-CCSD 
with the $\lambda$-MR-CCSD and the CAS-SD values. Results are also given using
the reference-independent dressing of the CAS-SD CI matrix. All the
applications are presented as energy differences with respect to the Full-CI
energy estimated by a CIPSI calculation with a second-order perturbative
correction. Figure~\ref{fig:npe} shows the difference of energy with respect to
the Full-CI along the reaction coordinate. Table~\ref{tab:npe} summarizes
the non-parallelism errors (NPE) and the maximum of the error obtained
along the curve.
The MR-CC treatment reduces the average and maximum error of the CAS-SD with
respect to Full-CI by a factor close to 4. The correction is larger when the
system involves an important number of inactive electrons (F$_2$, C$_2$H$_6$) than when
this number is small (BeH$_2$, N$_2$). One actually knows that the size-consistency
error of the CAS-SD treatment increases with the number of inactive
electrons, this error disappears in the MRCC treatment, which essentially
misses some fourth-order connected effects of the Triples. 

\begin{table*}
\caption{Non-parallelism errors (NPE) and maximum errors with respect to the Full-CI potential energy surface (m$E_{\rm h}$)}
\label{tab:npe}
\begin{tabular}{lcccccccc}
\hline
                              & \multicolumn{2}{c}{CAS-SD}    & \multicolumn{2}{c}{$\lambda$-MR-CCSD}    & \multicolumn{2}{c}{Dressed    CAS-SD} & \multicolumn{2}{c}{$\mu$-MR-CCSD}     \\ 
                              &     NPE      &   Max Error    &     NPE      &         Max Error         &     NPE      &       Max Error        &     NPE      &       Max Error        \\ 
\hline
          C$_2$H$_6$          &     5.1      &     35.5       &     3.6      &           8.4             &     1.3      &         8.3            &     3.5      &         8.3            \\ 
            F$_2$             &     3.8      &     19.8       &     3.1      &           4.0             &     0.9      &         4.2            &     1.6      &         3.9            \\ 
       C$_2$H$_4$ twist       &     1.5      &     27.7       &     0.3      &           6.7             &     0.7      &         7.2            &     0.5      &         6.7            \\ 
           BeH$_2$            &     2.9      &      4.1       &     1.3      &           1.8             &     2.0      &         2.4            &     1.8      &         2.2            \\ 
            H$_2$O            &     1.8      &      4.6       &     0.7      &           1.2             &     0.2      &         1.3            &     0.5      &         1.2            \\ 
      C$_2$H$_4$ stretch      &     2.8      &     22.1       &     1.6      &           5.3             &     1.7      &         6.2            &     1.6      &         5.2            \\ 
            N$_2$             &     1.8      &      9.0       &     1.0      &           2.2             &     1.7      &         3.9            &     0.7      &         2.8            \\ 
\hline
 F$_2$ $^3\Sigma_u^+ (m_s=1)$ &     2.5      &     18.6       &     1.3      &           3.3             &     1.3      &         3.5            &     1.2      &         3.3            \\ 
 F$_2$ $^3\Sigma_u^+ (m_s=0)$ &     2.5      &     18.6       &     1.2      &           1.8             &     1.3      &         3.5            &     1.1      &         3.3            \\ 
      HF (ground state)       &     2.6      &     14.6       &     1.8      &           3.4             &     2.1      &         4.4            &     1.8      &         4.1            \\ 
      HF (excited state)      &     3.3      &     20.9       &     8.8      &           8.5             &     10.5     &        10.1            &     7.1      &         8.3            \\ 
\hline
        F$_2$ (local)         &     3.8      &     19.8       &     1.2      &           3.2             &     1.5      &         3.1            &     1.0      &         4.6            \\ 
        N$_2$ (local)         &     1.8      &      9.0       &     3.8      &           5.0             &     1.1      &         3.5            &     1.1      &         2.8            \\ 
\hline
\end{tabular}
\end{table*}

\subsubsection*{Single-bond breaking}

We present here the single bond breaking of the $\sigma$ bonds of C$_2$H$_6$ and F$_2$ molecules
and of the $\pi$ bond of ethylene.
The active spaces were chosen with two electrons in two MOs, the minimum wavefunctions
to describe properly the dissociation of the molecules. 
In the case of ethane, the NPE of the CAS-SD is 5.1~m$E_{\rm h}$, and is reduced to 3.5\meh
with the $\mu$-MR-CCSD. The curve of the dressed CAS-SD has 
the lowest NPE (1.3~m$E_{\rm h}$). The curves obtained by both MR-CCSD methods
give equivalent results, with NPEs of 3.5 and 3.6\mehp

In the case of F$_2$ the NPE of the dressed CAS-SD is 0.9\meh and the NPE of
the $\mu$-MR-CCSD is 1.6~m$E_{\rm h}$, both better than the NPE of the
$\lambda$-MR-CCSD which has an NPE of 3.1\mehp Also, one can remark
here some numerical instabilities in the $\lambda$-MR-CCSD where the
curve is not perfectly smooth.

In the next example, the $\pi$ bond of ethylene is broken by the rotation of the
CH$_2$ fragments. The CAS-SD has an NPE of 1.5\mehc and using the dressed CAS-SD
reduces the NPE to 0.7\mehp The $\mu$-MR-CCSD gives an NPEs of 0.5\mehc
and the NPE obtained with the $\lambda$-MR-CCSD is slightly better with an NPE
of 0.3\mehp

\subsubsection*{Insertion of Be in H$_2$}

We present the results obtained by the insertion of a beryllium atom into the
H$_2$ molecule, which is a popular benchmark for MR-CC
methods.
The reference is still a CAS(2,2) for comparison with the literature, even
though this choice of reference is not the most appropriate for a correct
description of the reaction. The geometries are given by the relation
\begin{equation}
 \label{beh2_geom}
 z = 2.54 - 0.46 x \quad \text{(a.u.)}
\end{equation}
where the beryllium atom is at the origin and the hydrogen atoms are at the
coordinates $(x,0,\pm z)$. In this particular case, the $\mu$-MR-CCSD
gives a NPE of 1.8\meh which is larger than the NPE of 1.3\meh obtained
by the $\lambda$-MR-CCSD. This is due to only one point of the curve, the maximum 
which is higher by 0.4\mehc all the other points being very close by less
than 0.1\mehp Here, the dressed CAS-SD and the $\mu$-MR-CCSD are equivalent.

\subsubsection*{Two bond breaking}

For breaking two bonds we have used CAS(4,4) wave functions as the reference space.
The first example is the simultaneous breaking of the two O---H bonds of the
water molecule by stretching. Here, the CAS-SD exhibits a NPE of 1.8\meh which
is significantly improved to 0.2\meh with the dressed CAS-SD. 
The $\mu$-MR-CCSD, with an NPE of 0.5\mehc is slightly more parallel to the
Full-CI curve than the $\lambda$-MR-CCSD which has an NPE of 0.7\mehp

The second example is the double-bond breaking of ethylene by stretching. One
should first clarify that the energy differences in the figure do not match
those of the torsion along the bond because in the former example the reference
was a CAS(2,2), and here it is a CAS(4,4).
Dressing the CAS-SD reduces the NPE from 2.8\meh to 1.7\mehp One can remark a
discontinuity in the curve at large distances. 
The $\mu$-MR-CCSD and $\lambda$-MR-CCSD slightly improve the NPE to a value of 1.6\mehc
and both variants of the MR-CCSD are equivalent with smooth curves.

\subsubsection*{Triple-bond breaking}

N$_2$ is the typical benchmark for breaking a triple bond.
Here, we have used a CAS(6,6) reference wave function. At the CAS-SD level, the
NPE is 1.7\mehc and the dressed CAS-SD doesn't reduce the NPE.
Here, it is necessary to use reference-dependant amplitudes to
recover a low NPE~: 1.0\meh with the $\lambda$-MR-CCSD, and 0.7\meh
with the $\mu$-MR-CCSD.

\subsection{Excited states}
\label{section3c}

\subsubsection*{Triplet state of F$_2$}

\begin{figure}
\begin{center}
\includegraphics[angle=270,width=0.48\textwidth]{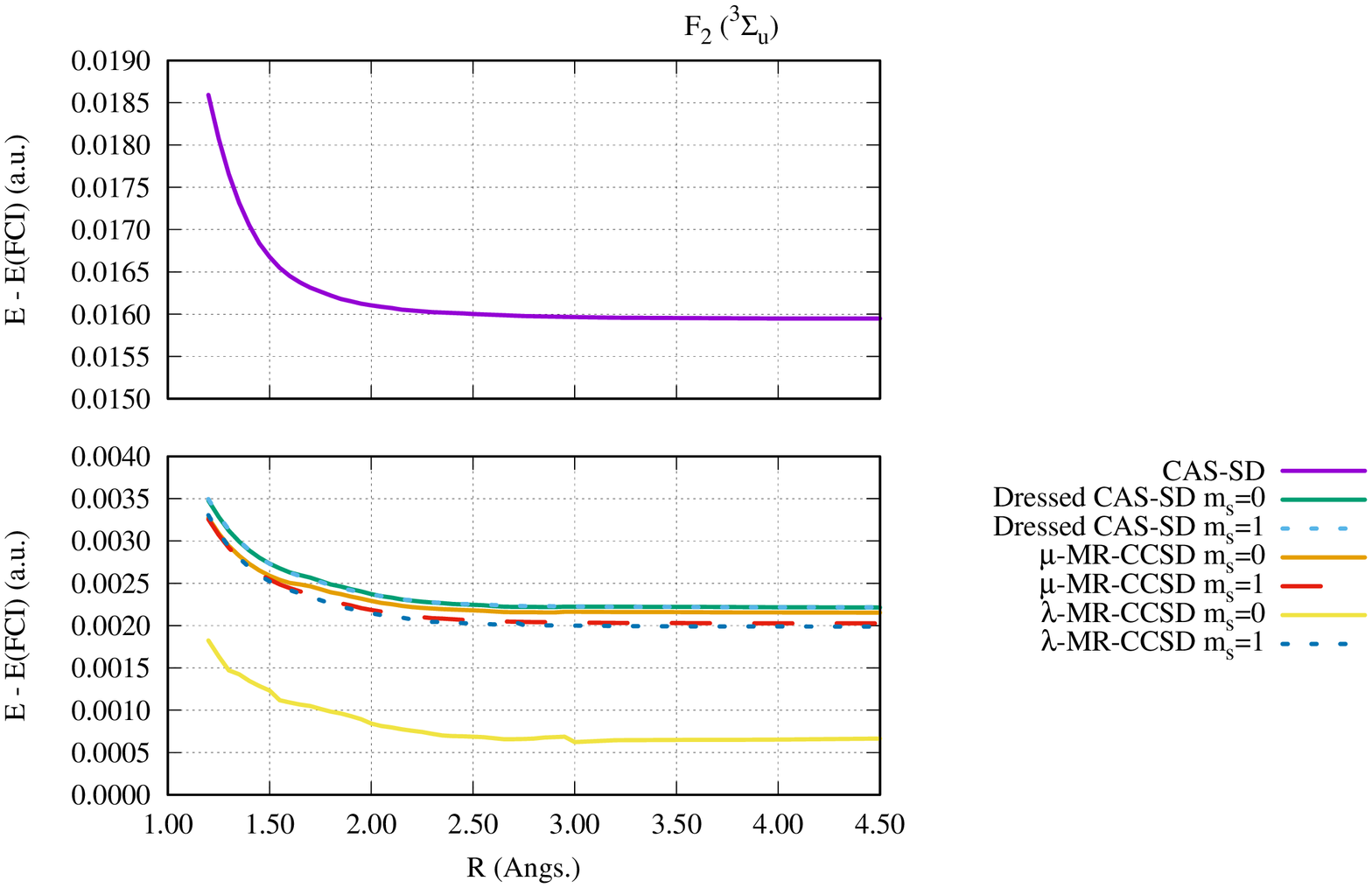} \\
\includegraphics[angle=270,width=0.48\textwidth]{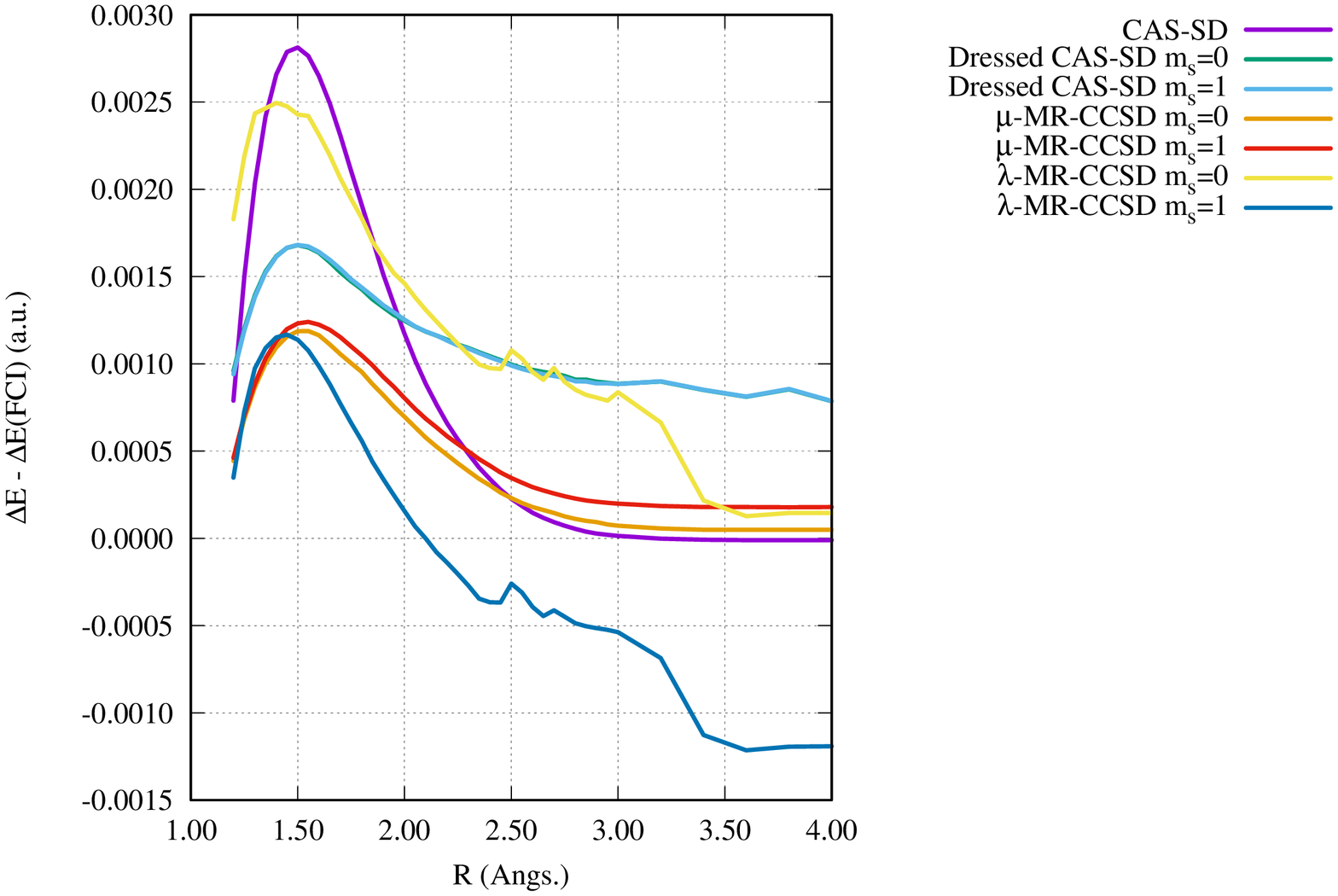}
\end{center}
\caption{F$_2$ $^3\Sigma_u^+$.  Difference with respect
to the Full-CI energy for the $m_s=0$ and $m_s=1$ wave functions (top), and
error on the singlet-triplet gap $\Delta E = E(^3\Sigma_u^+) - E(^1\Sigma_g^+)$ (bottom).
On both graphics, the two curves of the dressed CAS-SD coincide.}
\label{fig:f2_3}
\end{figure}

We report here calculations on the triplet state $^3\Sigma_u^+$ of F$_2$.  The
reference wave function was prepared in two different ways, both using
restricted open-shell Hartree-Fock molecular orbitals. The first reference
wave function labeled $m_s=1$ is a single open-shell determinant,
and the second wave function is the triplet $m_s=0$, made of two determinants
$1/\sqrt{2}(\alpha\beta - \beta\alpha)$. 

To ensure that the CAS-SD is a strict eigenfunction of the $\hat{S^2}$ operator, we have included
in $\Psi_{\rm SD}$ all the determinants with the same space part as the Singles
and Doubles with respect to the CAS.
These determinants are treated in the same way as Singles and Doubles and are
treated variationally in the diagonalizations. Of course, those which are Triples
or Quadruples with respect to $\Psi_{\rm ref}$ are excluded from 
the set of the $\{\alpha\}$ and have no effect in the dressing.

%At the CAS-SD level, the wave
%functions are equivalent and give the exact same energy, but it is interesting to
%see how the MR-CCSD and the dressed CAS-SD behave with those two wave
%functions (figure~\ref{fig:f2_3}). 

To reduce the computational cost, the Triples and Quadruples were not augmented with all the determinants
with the same space part. The absence of some determinants 
gives rise to a slight deviation ($< 10^{-6}$ a.u.) of $\langle \hat{S^2}
\rangle$ from the desired eigenvalue, and it is expected to have some impact on
the iterative dressing.
%If one considers the
%various determinants $\Phi_l^\nu$ of a given space part, characterized by the
%symbol $\nu$ and various spin distributions $k$, if one imposes the condition
%\begin{equation}
% \frac{c_k^\nu}{c_l^\nu} = \frac{\langle \Phi_k^\nu | \hat{H} | \Psi_0 \rangle}{\langle \Phi_l^\nu | \hat{H} | \Psi_0 \rangle}
%\end{equation}
%the vector 
%\begin{equation}
%|\Psi^\nu \rangle = \sum_k c_k^\nu | \Phi_k^\nu \rangle
%\end{equation}
%is an eigenfunction of $\hat{S^2}$ since $\Psi_0$ is an eigenfunction of  $\hat{S^2}$ and since
%$\hat{H}$ and $\hat{S^2}$ commute.
%This spin-decontaminated wave function,
%\begin{equation}
%|\Psi_{\rm CAS-SD}^\prime \rangle = \sum_I C_I |I\rangle + \sum_\nu c_\nu |\Psi_\nu\rangle 
%\end{equation}
%which is extremely close to $\Psi_{\rm CAS-SD}$ might be handled instead of
%$\Psi_{\rm CAS-SD}$ to generate the excitation operators  according to the
%procedures of ref.\cite{Giner_2016} or in the present work. These excitation
%operators would commute with $\hat{S^2}$, and so would be their products.  Then
%the determinants of the Triples and Quadruples would enter in eigenfunctions of
%$\hat{S^2}$ and the dressing would maintain this property. For simplicity in
%the implementation, which is determinant driven, we have not introduced this
%constraint, and
It is worth checking the effect of this deviation from the
exact spin multiplicity. The first test concerns the comparison of the $m_s=0$ and
$m_s=1$ components of a triplet state.

In all the cases, the NPE of the CAS-SD (2.6~m$E_{\rm h}$) is improved to a value of 
1.1--1.5\mehp As expected the two variants of the MR-CCSD are strictly
equivalent for $m_s=1$. Indeed, for both variants the usual single-reference
amplitudes $c_i/c_0$ are recovered.
The amplitudes of the $\lambda$-MR-CCSD lower the curve by 1\meh when going
from $m_s=1$ to $m_s=0$.
The dressed CAS-SD also gives a lower energy, but only by
0.5\mehp This is due to the increased number of degrees of freedom in the
fit of the amplitudes as no additional constraint is imposed to enforce the
$m_s$-invariance. But when the reference-dependence is introduced via the
$\mu_i$, it is imposed to recover the CAS-SD wave function which is $m_s$-invariant,
and this step compensates the additional freedom gained in the fitting, and
the $m_s=1$ and $m_s=0$ MR-CCSD curves differ by less than 0.1\mehp

If one considers the error on the singlet-triplet gap with respect to the Full-CI
reference, it appears clearly that the $\mu$-MR-CCSD gives the most
accurate results, with errors lying between 0.1\meh and 1.3\meh along the curve.

\subsubsection*{Avoided crossing in HF and LiF}

\begin{figure}
\begin{center}
\includegraphics[angle=270,width=0.9\columnwidth]{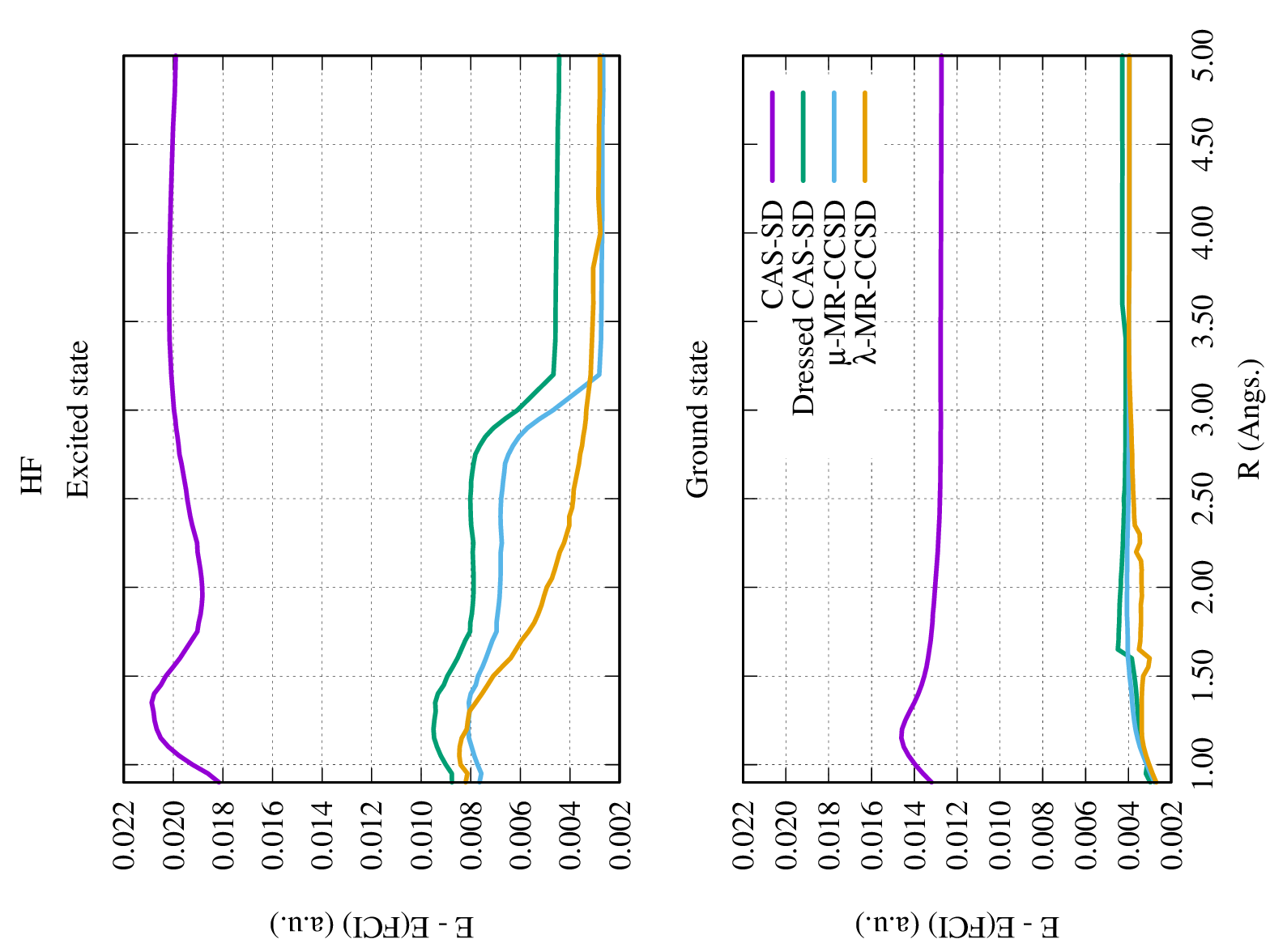}
\end{center}
\caption{Difference with respect to the Full-CI energy for the two lowest
$^1\Sigma^+$ states of HF.}
\label{fig:hf}
\end{figure}

\begin{figure}
\begin{center}
\includegraphics[angle=270,width=0.9\columnwidth]{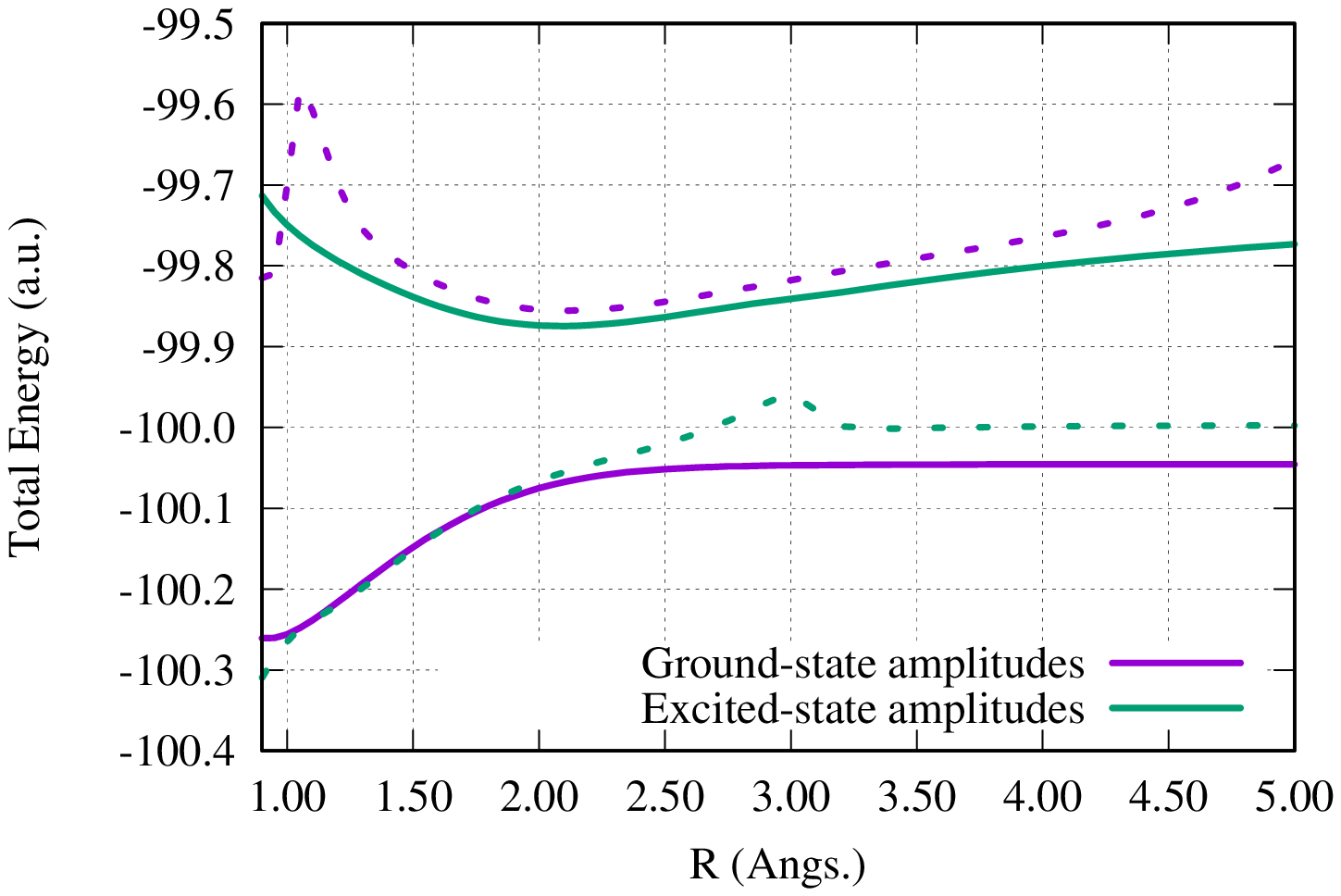}
\end{center}
\caption{Potential energy surfaces of the two lowest $^1\Sigma^+$ states of HF
with the $\mu$-MR-CCSD method.  The energy of the state corresponding to the
dressing is plotted in plain curves, and the energy of the other state is
plotted in dashed curves.}
\label{fig:hf_roots}
\end{figure}

We have calculated the potential energy surfaces of the two lowest $^1\Sigma^+$
states of HF, using as reference wave function the CAS(2,2) with state-averaged
CAS-SCF molecular orbitals in the aug-cc-pVDZ basis set.
Figure~\ref{fig:hf} shows the NPEs of the ground and excited states.
In the ground state, the NPE is 1.8\meh for both MR-CCSD variants, but
the $\lambda$-MR-CCSD shows some numerical instabilities, as opposed to the 
$\mu$-MR-CCSD which gives a very smooth curve.

In the excited state, the situation is different~: surprisingly the best NPE is obtained by
the CAS-SD, but this may be due to the fact the molecule is particularly favorable to
the CAS-SD : both H and H$^+$ have no correlation energy, so the wave functions of the
dissociated molecule can be expressed as a product.
The two variants of the MR-CCSD agree at short and long distances, but they differ
significantly between 2 and 3.0~\AA, after the region of the avoided crossing. To understand
these differences, we have plotted the two eigenvalues of the two state-specific Hamiltonians,
one dressed for the ground state and one dressed for the excited state.
It appears that between 2 and 3.0~\AA,  the lowest eigenvalue of the Hamiltonian dressed for the
excited state is very badly described. The reason is that the fitting procedure for the
amplitudes is a least-squares fit on the CAS-SD wave function of the state of interest, so
the quality of the dressing for the determinants which have small coefficients
on the state of interest but large coefficients on the other state will be very low.
The $\lambda$-MR-CCSD has amplitudes which depend less on the wave function,
so the quality is equivalent on both states, and the choice of these amplitudes
is better suited for calculating excited states within the same symmetry. 

\begin{figure}
\begin{center}
\includegraphics[angle=270,width=0.9\columnwidth]{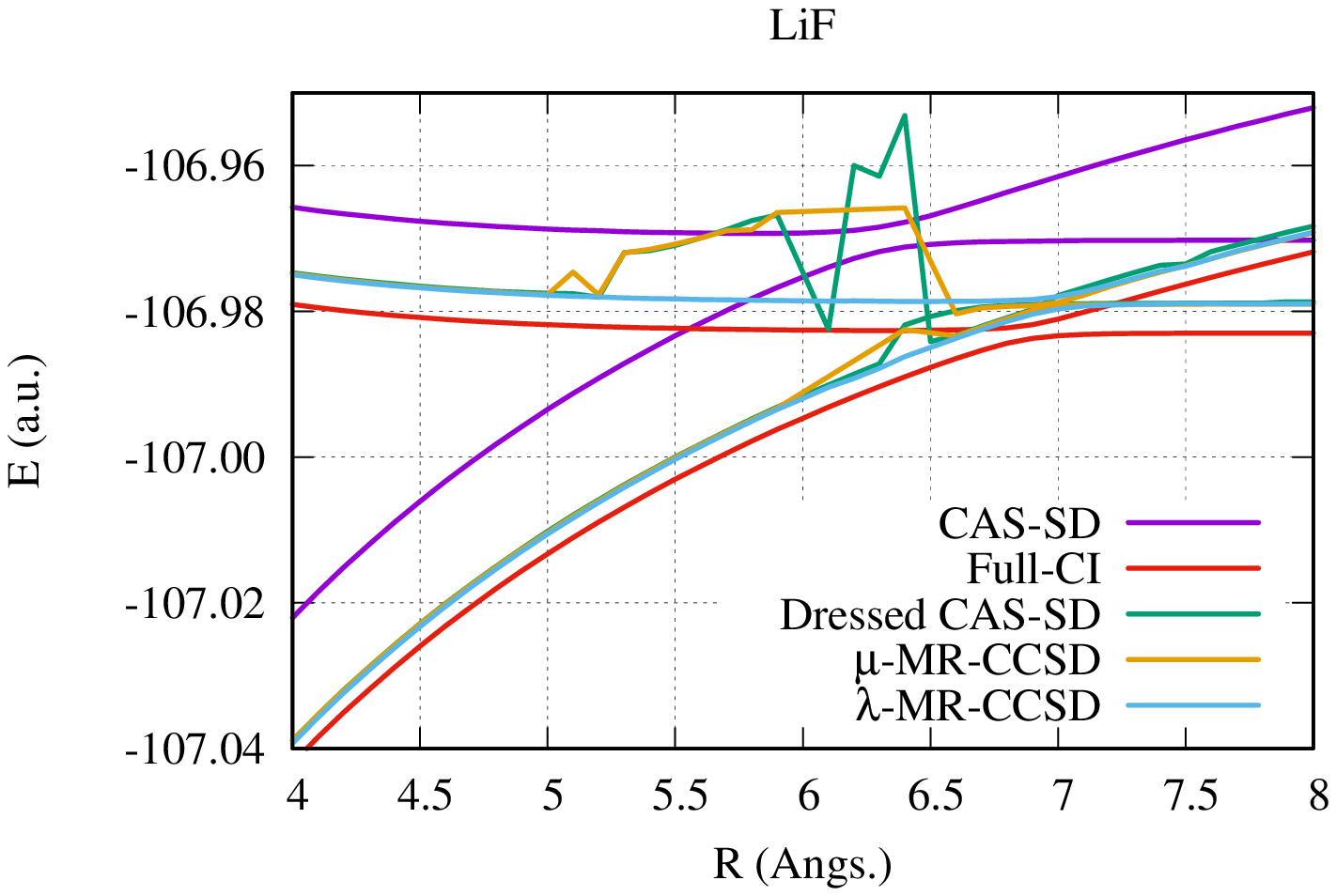}
\end{center}
\caption{Potential energy surfaces of the two lowest $^1\Sigma^+$ states of LiF.}
\label{fig:hf_roots}
\end{figure}
In figure~\ref{fig:hf_roots} we have represented the avoided crossing of LiF, also
calculated with the aug-cc-pVDZ basis set.
The physical situation is similar to HF, but the energy difference between the
ground and the excited states is much smaller.
A striking result is that the $\lambda$-MR-CCSD, although being state-specific, is
able to reproduce very well the whole potential energy surfaces of both states.
The position of the avoided crossing is very well reproduced by
the three methods~: the CAS-SD crosses at 6.3~\AA, the Full-CI crosses at 6.8~\AA and
the dressed CAS-SD and the two MR-CCSD variants cross at 6.9~\AA. The $\mu$-MR-CCSD
and $\lambda$-MR-CCSD coincide in the short-range ($\le 5~\AA$) and in the long
range ($\ge 7.2~\AA$), but when the two states become very close in energy in
the region of the crossing the dressed CAS-SD and the $\mu$-MR-CCSD are
unable to give sensible values. This disappointing result motivates a future
work on a multi-state $\mu$-MR-CCSD.

\subsection*{Sensitivity to the choice Molecular Orbitals}

\begin{figure}
\begin{center}
\includegraphics[angle=270,width=0.9\columnwidth]{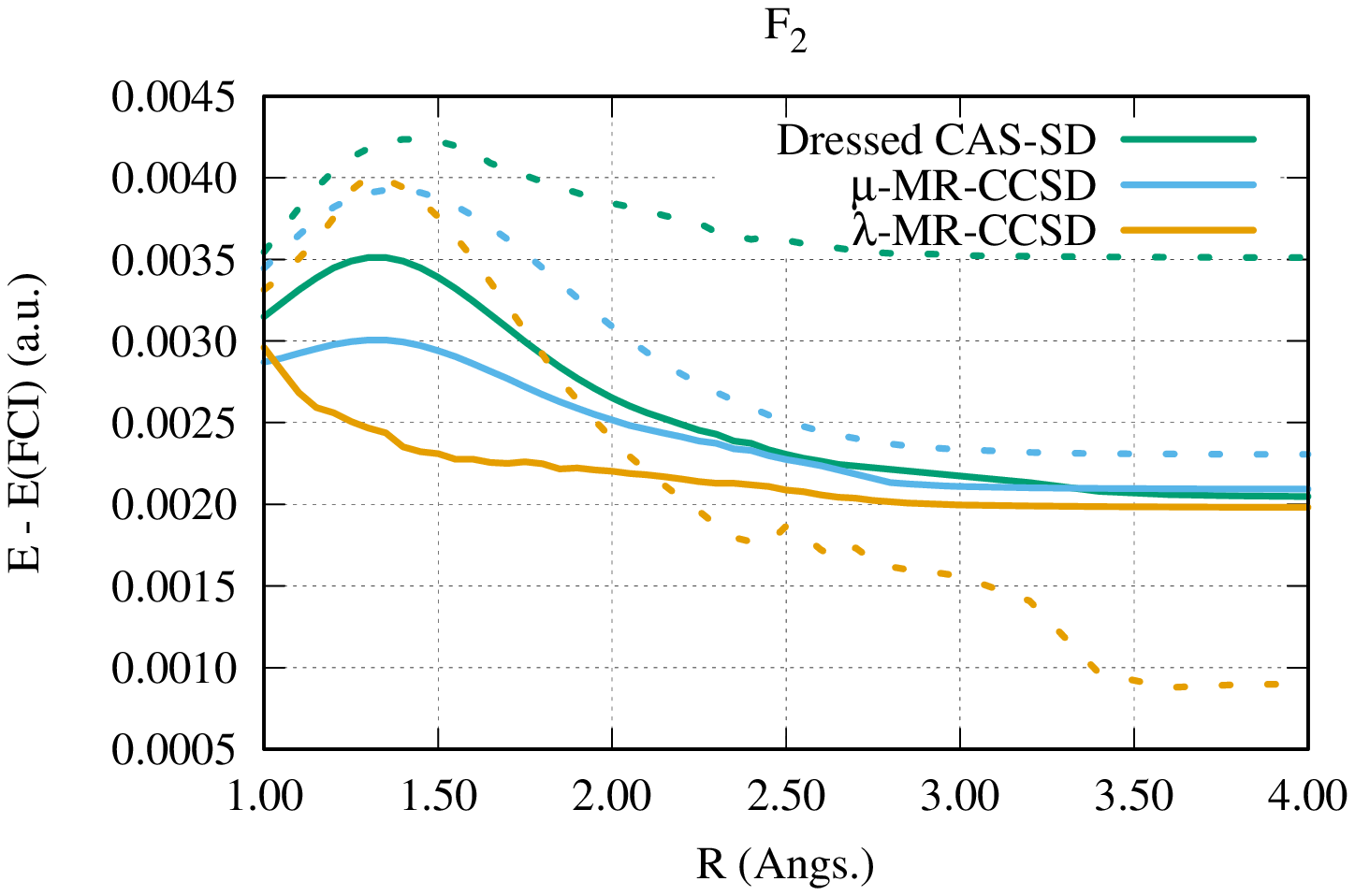}
\includegraphics[angle=270,width=0.9\columnwidth]{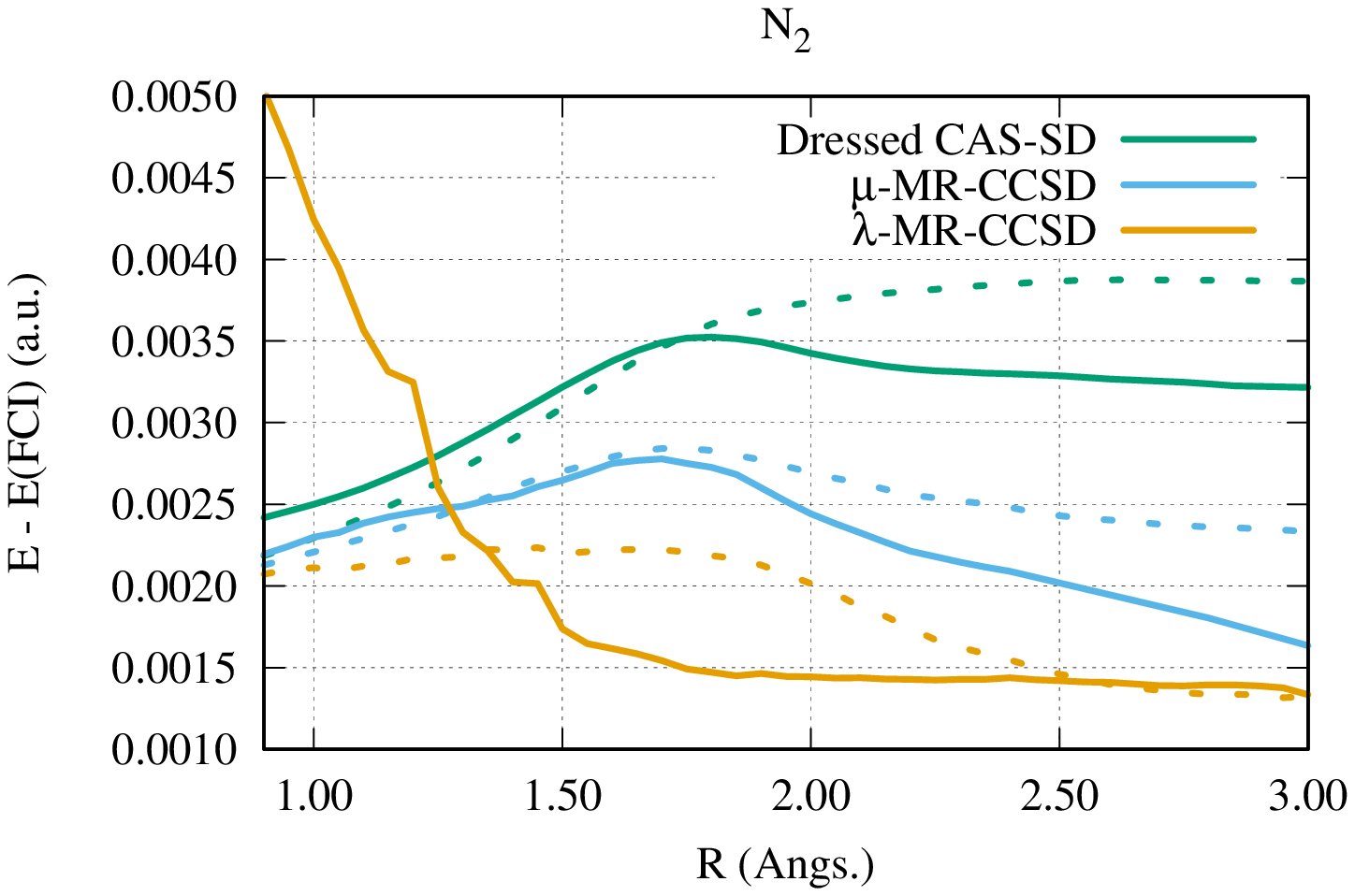}
\end{center}
\caption{Comparison between pseudo-canonical (dashed curves) and
localized (plain curves) MOs in F$_2$ and N$_2$. Difference with respect to the Full-CI
energy.}
\label{fig:local}
\end{figure}

The $\mu$-MR-CCSD algorithm we propose is in the Jeziorski-Monkhorst framework, so it
is not invariant with respect to the choice of molecular orbitals. In this
section, we checked its sensitivity to the
choice of the MO set by comparing results obtained with pseudo-canonical
CAS-SCF orbitals and with localized MOs in the F$_2$ and N$_2$ molecules
(figure~\ref{fig:local}). 

%Dans la discussion sur cette sensibilité quand tu passes à N2, on a oublié le passagee local à canonique énoncé à propos de F2.

In the F$_2$ molecule, using localized MOs is a better choice than
the pseudo-canonical MOs. The best NPE is obtained by the $\mu$-MR-CCSD
method with a value of 1.1\mehp
In the case of N$_2$, the situation is different~: the NPE of the
$\lambda$-MR-CCSD goes from 0.9\meh to 3.7\mehc and the
NPE of the $\mu$-MR-CCSD increases from 0.7\meh to 1.1\mehp 
On the other hand, the dressed CAS-SD gives a better NPE with local
orbitals, going from 1.7\meh to 1.1\mehp

The fact that the $\mu$-MR-CCSD is less sensitive to the MO set than the
$\lambda$-MR-CCSD can be understood.
By changing the MO set, a single excitation rotates into a combination
of single and double excitations. In the $\lambda$-MR-CCSD method, the amplitudes are
calculated by taking into account the matrix elements of the Hamiltonian,
which are of different nature depending on the degree of excitation,
so the amplitudes are expected to change significantly.
In the $\mu$-MR-CCSD variant, the amplitudes are adjusted in such a way that they
fit the CAS-SD wave function, which is invariant by rotation of the MOs. Therefore,
it is expected to be more robust with respect to the MO set.

\section{Conclusions}

We have proposed a method to determine reference-independent amplitudes by fitting
the CAS-SD CI vector. These amplitudes may be used to perform a state-specific
iterative dressing of the CAS-SD Hamiltonian in order to take into account the
effect of the Triples and Quadruples in the spirit of the Coupled Cluster
formalism. Alternatively, these amplitudes may be rescaled to reproduce the exact
coefficients of the singles and doubles
to introduce a reference-dependent character. In that case, the CAS-SD CI
vector is recovered by the application of $(1+\hat{T})$ on the reference wave
function, so we reach here the Jeziorski-Monkhorst Coupled Cluster formalism.

The CAS-SD dressed with reference-independent amplitudes gives excellent results for
single-bond breaking (F$_2$ and ethane) and the simultaneous breaking of the two
O---H bonds of water, with a non-parallelism error lower than the milli-Hartree.
When the active space becomes larger, it is necessary to go to the
reference-dependent MR-CCSD introducing the $\mu$ factors in Eq.~\ref{eq:mu}.
In the case of ethylene and N$_2$, this keeps the NPE to a value close to
the milli-Hartree.

We have shown numerically that the here-proposed amplitudes are not very
sensitive to the value of $m_s$ for open-shell systems, and to the choice of
the molecular orbitals. This is clearly an improvement compared the 
amplitudes proposed earlier\cite{Giner_2016}.
But we have also shown that the former amplitudes are a better choice
when computing excited states of the same symmetry because the here-proposed
amplitudes have a much more pronounced state-specific character which may be
disadvantageous if the states are too close in energy. This problem can be cured
by leaving the state-specific formalism for a multi-state formalism\cite{Malrieu_2013},
and this will be the object of a future work.

{\it Acknowledgments.}
This work has been made through generous computational support from CALMIP
(Toulouse) under the allocation 2015-0510, and GENCI under the allocation
x2015081738.

\bibliography{MRCC}
\end{document}